\documentstyle[12pt,epsf]{article}

\begin{document}

\newcommand {\bea}{\begin{eqnarray}}
\newcommand {\eea}{\end{eqnarray}}
\newcommand {\be}{\begin{equation}}
\newcommand {\ee}{\end{equation}}

\def\IR{{\hbox{{\rm I}\kern-.2em\hbox{\rm R}}}}
\def\IH{{\hbox{{\rm I}\kern-.2em\hbox{\rm H}}}}
\def\IC{{\ \hbox{{\rm I}\kern-.6em\hbox{\bf C}}}}
\def\IZ{{\hbox{{\rm Z}\kern-.4em\hbox{\rm Z}}}}

\begin{titlepage}
\rightline{EFI-99-42}

\rightline{hep-th/9909102}

\vskip 3cm
\centerline{\Large{\bf Rotating Kaluza-Klein Black Holes}}
\vskip 2cm
\centerline{Finn Larsen}
\vskip 12pt
\centerline{\sl the Enrico Fermi Institute}
\centerline{\sl University of Chicago}
\centerline{\sl 5640 S. Ellis Ave., Chicago, IL 60637, USA}
\centerline{\texttt{flarsen@theory.uchicago.edu}} 
\vskip 2cm

\begin{abstract}
All regular four-dimensional black holes are constructed in the theory 
obtained by Kaluza-Klein reduction of five-dimensional Einstein gravity. 
They are interpreted in string theory as rotating bound states of D0- and 
D6-branes. In the extremal limit the solutions are stable, due to angular 
momentum conservation. The thermodynamics, the duality symmetries, 
and the near-horizon limit are explored.
\end{abstract}
\end{titlepage}

\newpage

\section{Introduction}
Black holes in Kaluza-Klein theories have attracted tremendous attention 
over the last decades. A driving force of this interest is their role as 
low-energy approximations to string theory. The simplest Kaluza-Klein 
theory is obtained by starting with pure Einstein gravity in five dimensions,
and dimensionally reduce to four dimensions. The field content of the
resulting theory is gravity, the dilaton $\Phi_{4}$, and the gauge field 
$A_{\mu}$; the Lagrangian is:
\be
L = {1\over 16\pi G_{4}}\int d^{4}x \left[R-
2\partial_{\mu}\Phi_{4}\partial^{\mu}\Phi_{4} 
-{1\over 4}e^{-2\sqrt{3}\Phi_{4}}F_{\mu\nu}F^{\mu\nu} \right]~,
\label{eq:lag}
\ee
in the Einstein frame. The most general stationary black hole solutions 
to this theory are parametrized by the mass $M$, the angular momentum $J$, 
and the electric/magnetic charges of the gauge field, $Q/P$. The purpose of 
this paper is to construct this four-parameter family of black holes.

The electromagnetic field created by electric and magnetic sources, charged 
with respect to the same $U(1)$ field, carry angular momentum 
satisfying the universal bound:
\be
J \geq {n_{Q}n_{P}\over 2}~,
\label{eq:jbound1}
\ee
where $n_{Q,P}$ are the quantized values of the charges. Thus the properties 
of dyonic black holes are intimitely related to rotation. It is ultimately 
this feature which makes it interesting to construct the full family of
black holes, including angular momentum. 

As the spin of a black hole is increased the inner and outer horizons 
approach each other, eventually merging, and then exhibiting a naked 
singularity. Very rapidly spinning black holes are thus singular. 
The precise regularity bound on the angular momentum will be determined 
in the course of the present investigation. It is:
\be
J < {n_{Q}n_{P}\over 2}~,
\label{eq:jbound2}
\ee
in the extremal limit. This is an interesting result, in view of 
(\ref{eq:jbound1}). It shows that the regular black holes are precisely 
those which {\it cannot} form from widely separated sources carrying
electric and magnetic charges. Stated another way, the 
regular black holes are prevented from decaying into constituent parts 
by angular momentum conservation. This is significant because the 
mass of the black hole exceeds the sum of its constituent masses;
so energy conservation is not sufficient to stabilize the black holes 
against spontaneous fragmentation. 

The black hole is thus stable in the extremal limit; so it is reasonable 
to describe it as the ground state in a conformal field theory. Moving 
away from extremality, 
the black holes exhibit thermal properties, as expected. These are interpreted 
in terms of perturbations of the conformal field theory and should remain 
under control, as long as they are small. The prospects of a precise microscopic 
description of the black holes, even though they are not supersymmetric, is 
the ultimate goal of the inquiry. However, this will be pursued elsewhere; 
the discussion in this paper focusses on the classical properties of the 
black holes.

The present investigation is motivated by several relations to string theory. For 
example, add six additional toroidally compactified dimensions, and interpret 
the compact Kaluza-Klein direction as the M-theory circle. Then the solutions 
are charged with respect to the ``electric'' charge of $D0$-branes and
the ``magnetic'' charge of $D6$-branes, fully wrapping the six inert dimensions.
Thus the solutions can be interpreted as rotating bound states of 
$D0$- and $D6$-branes. As discussed above, angular momentum conservation 
implies that such bound states are stable, even though they are not 
supersymmetric. The argument may explain the stability to the leading
order noted in the string theory description of the 
$D0-D6$ system~\cite{Taylor:1997ay}.

For another application, recall that general four dimensional black holes 
in $N=4,8$ string theory are generated by black holes depending on 
5 parametric charges~\cite{Cvetic:1996zq}. By now it is standard to 
consider four of these charges~\cite{Cvetic:1996uj,Horowitz:1996ac}, 
but the fifth charge is difficult and our understanding is incomplete, even 
at the level of classical solutions; for discussion see 
{\it e.g}~\cite{Cvetic:1997im}. The problematic fifth charge parametrizes 
the inner product of the electric and magnetic charge vectors; when it is 
absent, the four charges refer to independent $U(1)$'s, up to duality. In 
the present work there is only one $U(1)$ field, so electric and magnetic 
charges are necessarilly parallel. In a sense, there is {\it only} the fifth 
charge, and therefore good opportunity to study its properties.

The paper is organized as follows. In section~\ref{gen}, we review the
solution generating technique, following Sen~\cite{Sen:1995eb}. The resulting 
solution is presented in its five-dimensional form, along with necessary 
notation. There is also a discussion of various special cases, and the 
relation to previously known solutions. Subsequently, in section~\ref{bh}, 
the properties of the corresponding black holes in four dimensions are 
derived. The various subsections are quasi-independent, each 
considering {\it e.g.} duality symmetries, 
thermodynamic properties, dipole moments, or
the near-horizon geometry.

\section{The Solution}
\label{gen}
In this section we review the solution generating technique, following
Sen~\cite{Sen:1995wr,Sen:1995eb}. This leads to the solution, of course; 
but it also serves to describe a specific embedding into string 
theory, and to explain how a potential Taub-NUT singularity is avoided.
Sen considered the heterotic string theory but for the present purposes it 
is sufficient to consider the five dimensional vacuum sector. This truncation 
simplifies the problem and further gives a universal embedding, valid for all
the string theories.

\subsection{The Solution Generating Technique}
The idea of the solution generating technique is to note that, being 
interested in stationary solutions, time can be assumed compact,
for the purpose of the equations of motions and their symmetries.
This procedure yields an effective theory in three dimensions. 
In the present context the compactification from five to three dimensions 
gives an $SO(2,2)$ T-duality symmetry, with transformations acting as 
conjugations on the moduli matrix:
\be
M = \pmatrix{
 G^{-1} & G^{-1}B \cr
 -BG^{-1} & G}~,
 \label{eq:Mmatrix}
\ee
where $G$ and $B$ are $2\times 2$ matrices with t/y indices. The 
labels in the explicit computations below are chosen so that
$M={\rm diag}[g_{yy}^{-1},g_{yy},g_{tt}^{-1},g_{tt}]$ for diagonal metrics.

The components of the metric and the B-field having one azimuthal index, and 
the other in the t/y directions, correspond to magnetic fields in three 
dimensions. They are represented as four pseudoscalar potentials $\Psi$ 
after the dualization:
\be
 -\sqrt{g_{3}} e^{-2\Phi_{3}} g^{\mu\mu^{\prime}}g^{\nu\nu^{\prime}}
 (ML)_{ab} F_{\mu^{\prime}\nu^{\prime}}^{(b)}
 = \epsilon^{\mu\nu\rho}\partial_{\rho}\Psi^{(a)}~.
\label{eq:psidef}
\ee
Here $L = {\rm diag}[\sigma_{1},\sigma_{1}]$, where:
\be
\sigma_{1}= \pmatrix{ 0 & 1 \cr
                      1 & 0 }~.
\ee
The four scalars fields $\Psi$ and the dilaton $\Phi_{3}$ are combined 
with the matrix $M$ to form an extended moduli matrix ${\cal M}$, of size 
$6\times 6$:
\be
{\cal M} = \pmatrix{
 M -e^{2\Phi_{3}}\Psi\Psi^{T} & e^{2\Phi_{3}}\Psi &M{\tilde\Psi} -{1\over 
 2}e^{2\Phi_{3}}\Psi~\Psi^{T}{\tilde\Psi} \cr 
 e^{2\Phi_{3}}\Psi & -e^{2\Phi_{3}}& -{1\over 
 2}e^{2\Phi_{3}}\Psi^{T}{\tilde\Psi}        \cr
 {\tilde\Psi}^{T}M +{1\over 
 2}e^{2\Phi_{3}}\Psi^{T}~\Psi^{T}{\tilde\Psi} & {1\over 
 2}e^{2\Psi_{3}}\Psi^{T}{\tilde\Psi} &  -e^{-2\Phi_{3}} + 
 {\tilde\Psi}^{T}M{\tilde\Psi}
 -{1\over 4}e^{2\Phi_{3}} (\Psi^{T}{\tilde\Psi})^{2}}
 \label{eq:mcaldef}
\ee 
where ${\tilde\Psi}\equiv L\Psi$. Note that ${\cal M}$ is symmetric
${\cal M} = {\cal M}^{T}$.

The group of solution generating transformations is formed by repeated 
application of T-duality and S-duality; the resulting group is $SO(3,3)$. 
The solution generating matrices $\Omega\in SO(3,3)$ satisfy:
\be
\Omega {\cal L}\Omega^{T} = {\cal L}~,
\label{eq:gendef}
\ee
where ${\cal L}= {\rm diag}[\sigma_{1},\sigma_{1},\sigma_{1}]$.
They leave the effective three dimensional metric invariant, and 
transforms the extended moduli matrix as:
\be
{\cal M}^{\prime} = \Omega {\cal M} \Omega^{T}~.
\ee

The solutions are required to have canonical moduli at infinity; {\it i.e.} 
the asymptotic moduli matrix must be 
${\cal M}_{\rm as} = {\rm diag}[I_{2}, -I_{2}, -I_{2}]$, where
$I_{2}$ is the $2\times 2$ identity matrix. This reduces
the group of solution generating transformations 
$SO(3,3)\to SO(2,1)\times SO(2,1)$. The unbroken group includes the 
compact $U(1)$ subgroup of the $SL(2,\IR)$ S-duality group, which leaves 
invariant the four dimensional solution used as starting point; so 
the space of solutions becomes a five-dimensional coset:
\be
\left( SO(2,1)\times SO(2,1)\right)/SO(2)~.
\ee
The five parameters are interpreted as electric/magnetic Kaluza-Klein charge, 
electric/magnetic B-field charge, and a Taub-NUT parameter. Removing
the charges associated with the B-field, {\it i.e.} the FS- and NS5-charges,
we are left with the diagonal $SO(2,1)$ of the coset above. 
Three generating elements of this group are:
\bea
\Omega_{E} &=& 
\pmatrix{
\cosh\alpha ~I_{2} & \sinh\alpha ~I_{2} & 0 \cr
\sinh\alpha ~I_{2} & \cosh\alpha ~I_{2} & 0 \cr
0 & 0 & I_{2}}~, 
\label{eq:omegae}\\
\Omega_{M} &=& \pmatrix{
\cosh\beta~I_{2} & 0 & \sinh\beta~\sigma_{1}  \cr
0 & I_{2} & 0 \cr
\sinh\beta~\sigma_{1} & 0& \cosh\beta~I_{2}}~,
\label{eq:omegam} \\
\Omega_{TN} &=& \pmatrix{ I_{2} & 0 & 0 \cr
0 &\cos\gamma~I_{2} & \sin\gamma~\sigma_{1}  \cr
0 & -\sin\gamma~\sigma_{1} & \cos\gamma~I_{2}  }~.
\label{eq:omegatn}
\eea
When acting on a background Schwarzchild or Kerr black hole, each of these 
transformations create a solution with the designated charge. 
The three infinitesimal generators satisfy the $SO(2,1)$ algebra, so 
together they generate a three parameter family of solutions, parametrized 
by the three charges.

In the present work only regular solutions are of interest; so the 
Taub-NUT charge must vanish. This can be cumbersome to achieve, because the 
transformations (\ref{eq:omegae}-\ref{eq:omegatn}) do not commute. For 
example, generating first a KK-monopole using $\Omega_{M}$, and then 
applying the boost $\Omega_{E}$, gives a solution with magnetic and elctric 
KK-charge, but also Taub-NUT charge. The standard remedy is to further
act with some $\Omega_{TN}$, with the parameter chosen to cancel the 
unwanted Taub-NUT charge. Solutions constructed this way tend to be quite unwieldy, 
due to the constraint of vanishing Taub-NUT charge. In particular, the 
symmetry between electric and magnetic charges is generally obscured. 
The strategy in the present construction is to solve the Taub-NUT constraint 
early on, implementing the result directly in the $SO(2,1)$ matrix. 
This leads to the two-parameter family of transformations:
\be
  \Omega=\pmatrix{
  \sqrt{qp\over 4m^{2}}I_{2}& \sqrt{(q-2m)(p+2m)\over 8m^{2}} 
  I_{2}& \sqrt{(q+2m)(p-2m)\over 8m^{2}}\sigma_{1}  \cr
  \sqrt{{p\over p+q}{q^{2}-4m^{2}\over 4m^{2}}}I_{2}& 
  \sqrt{{q\over p+q}{(q+2m)(p+2m)\over 8m^{2}}}I_{2}& 
    \sqrt{{q\over p+q}{(q-2m)(p-2m)\over 8m^{2}}} \sigma_{1} \cr 
      \sqrt{{q\over p+q}{p^{2}-4m^{2}\over 4m^{2}}}\sigma_{1}& 
  \sqrt{{p\over p+q}{(q-2m)(p-2m)\over 8m^{2}}} \sigma_{1}& 
    \sqrt{{p\over p+q}{(q+2m)(p+2m)\over 8m^{2}}} I_{2}} ~.
\ee
These matrices indeed belong to the $SO(2,1)$ of interest and, as shown below, 
further do not lead to Taub-NUT charge. When only electric or magnetic charges 
are present the correspondence with the parameters in $\Omega_{E}$ and 
$\Omega_{M}$ is $q = 2m\cosh^{2}\alpha$ and $p = 2m\cosh^{2}\beta$, 
respectively; but in general there is no simple relation to the 
familiar ``boost'' parameters.

\subsection{The Explicit Computation}
\label{comp}
We now turn to the explicit computation. The starting point is the 
standard Kerr black hole in four dimensions. In three 
dimensional form it is described by the metric:
\be
ds^{2}_{3}=H_{3} \left( {dr^{2}\over\Delta} + d\theta^{2} + 
{\Delta\over H_{3}}\sin^{2}\theta d\phi^{2} \right)~,
\ee
where:
\bea
H_{3} &=& r^{2}+a^{2}\cos^{2}\theta-2mr~,\\
\Delta &=& r^{2}+a^{2}-2mr~,
\eea
and the moduli:
\be
  {\cal M}=\pmatrix{
   1 & 0  & 0 & 0 & 0 & 0 \cr
   0 & 1  & 0 & 0 & 0 & 0 \cr
   0 & 0  & -f^{-1} & 0 & 0 & -g \cr
   0 & 0 & 0& -f(1+g^{2}) & g & 0 \cr
   0 & 0 & 0& g & -f^{-1} & 0 \cr
   0 & 0 & -g & 0 & 0 & -f(1+g^{2}) }~,
\ee
where:
\bea
f &=& {r^{2}+a^{2}\cos^{2}\theta - 2mr\over 
r^{2}+a^{2}\cos^{2}\theta} ~,\\
g &=& {2ma\cos\theta\over r^{2}+ a^{2}\cos^{2}\theta -2mr}~.
\eea
It is useful to note:
\bea
f(1+g^{2}) -1 &=& - ~2m(r-2m) H_{3}^{-1}~,\\
f^{-1} -1 &=& 2mr H_{3}^{-1}~.  
\eea

It follows without detailed computation from the structure of the various 
matrices that the components 
${\cal M}^{\prime}_{15}, {\cal M}^{\prime}_{35}, {\cal M}^{\prime}_{23}$ 
vanish identically. This implies that the fields carrying 
axion- , NS5- and FS-charges are consistently set to zero. 
More importantly, the leading term in ${\cal M}^{\prime}_{45}$ 
cancels, ensuring a vanishing Taub-NUT charge (the subleading terms 
are related to angular momentum.) An explicit computation 
further shows the relation:
\be
{\cal M}^{\prime}_{11}{\cal M}^{\prime}_{33}-
({\cal M}^{\prime}_{13})^{2} = {\cal M}^{\prime}_{55}~,
\ee
which implies that the five-dimensional dilaton $e^{2\Phi_{5}}=1$.
The remaining components of ${\cal M}^{\prime}$ can now be interpreted
consistently in terms of the five dimensional geometry. It can
be extracted from just a few elements of ${\cal M}^{\prime}$, {\it e.g}:
\bea
e^{2\Phi_{3}}&=& - {\cal M}^{\prime}_{55}~, \\
G_{tt}^{-1} &=& {\cal M}_{33}^{\prime}~,\\
A_{t} &=& - {{\cal M}_{13}^{\prime}\over {\cal M}^{\prime}_{33}}~,\\
G_{yy} & = & {{\cal M}^{\prime}_{33}\over {\cal M}^{\prime}_{55}}~,\\
\Psi^{(2)} & = & -{{\cal M}^{\prime}_{25}\over {\cal M}^{\prime}_{55}}~,\\
\Psi^{(4)} & = & -{{\cal M}^{\prime}_{45}\over {\cal M}^{\prime}_{55}}~.
\eea
The next step is to invert the definitions of $\Psi^{(2,4)}$ from 
(\ref{eq:psidef}),and find the corresponding gauge fields $A_{\phi}^{(1,3)}$. 
This is a lengthy computation. The $A_{\phi}^{(3)}$ gives the azimuthal 
part of the four-dimensional metric (denoted $B_{\phi}$ below). To find the 
azimuthal part of the gauge potential one must further transform from the 
effective three dimensional metric to the five dimensional form using:
\be
A_{\phi} = A_{\phi}^{(1)} + A_{t} A_{\phi}^{(3)}~.
\ee

\subsection{The Result}
\label{string}
The explicit computations lead to the five dimensional metric:
\be
ds^{2}_{5} = {H_{2}\over H_{1}}(dy + {\bf A})^{2}
-{H_{3}\over H_{2}}(dt + {\bf B})^{2}
+H_{1}({dr^{2}\over\Delta}+d\theta^{2}+{\Delta\over 
H_{3}}\sin^{2}\theta d\phi^{2})~,
\label{eq:string}
\ee
where:
\bea
H_{1} &=& r^{2}+a^{2}\cos^{2}\theta+r(p-2m)+{p\over 
p+q}{(p-2m)(q-2m)\over 2} -\nonumber \\ &~&\qquad - {p\over 2m(p+q)} 
\sqrt{(q^{2}-4m^{2})(p^{2}-4m^{2})}~a\cos\theta~,\\
H_{2} &=& r^{2}+a^{2}\cos^{2}\theta+r(q-2m)+{q\over 
p+q}{(p-2m)(q-2m)\over 2} +\nonumber \\ &~&\qquad + 
{q\over 2m(p+q)} 
\sqrt{(q^{2}-4m^{2})(p^{2}-4m^{2})}~a\cos\theta~,\\
H_{3} &=& r^{2}+a^{2}\cos^{2}\theta-2mr~,\\
\Delta &=& r^{2}+a^{2}-2mr~,
\eea
are quadratic functions of the Boyer-Lindquist type radial variable 
$r$ (we repeated $H_{3},\Delta$ for easy reference); and the 1-forms:
\bea
{\bf A} &=& -\left[ 2Q(r+{p-2m\over 2}) + 
\sqrt{q^{3}(p^{2}-4m^{2})\over 4m^{2}(p+q)}a\cos\theta\right] H_{2}^{-1} dt
\nonumber \\
&~&\qquad-\left[2P(H_{2}+ a^{2}\sin^{2}\theta)\cos\theta+
\sqrt{p(q^{2}-4m^{2})\over 
4m^{2}(p+q)^{3}}\times \right.
\nonumber \\ &~& 
\left.\phantom{\sqrt{p q^{2})\over 
4m^{2}}} \times
\left[(p+q)(pr-m(p-2m))+q(p^{2}-4m^{2})\right]a\sin^{2}\theta 
\right]H_{2}^{-1}d\phi~,
\label{eq:agauge} \\
{\bf B} &=& \sqrt{pq}{(pq+4m^{2})r-m(p-2m)(q-2m)\over 2m(p+q)H_{3}}~a 
\sin^{2}\theta d\phi~,
\label{eq:bgauge}
\eea
play the role of gauge potentials in the effective three-dimensional 
theory, obtained by compactifying $t$ as well as $y$. 
It is sometimes an advantage to write 
$A_{\phi}$ in the alternative form:
\bea
A_{\phi}&=& -\left[2P(\Delta + rq+
{q(p-2m)(q-2m)\over 2(p+q)})\cos\theta+
{2Pq\sqrt{(q^{2}-4m^{2})(p^{2}-4m^{2})}\over 2m(p+q)}
\right. \nonumber \\ &~&\qquad +\left. 
\sqrt{p(q^{2}-4m^{2})\over 4m^{2}(p+q)}(pr-m(p-2m))~a\sin^{2}\theta 
\right]H_{2}^{-1}~.
\eea

The four parameters $(m,a,q,p)$ appearing in the solution are related to 
the physical mass (M), angular momentum (J), electric charge (Q), and magnetic 
charge (P) through:
\bea
2G_{4}M&=& {p+q\over 2}~,
\label{eq:paraM} \\
G_{4}J &=& {\sqrt{pq}(pq+4m^{2})\over 4m(p+q)}~a~,
\label{eq:paraJ} \\
Q^{2} &=& {q(q^{2}-4m^{2})\over 4(p+q)}~,
\label{eq:paraQ} \\
P^{2} &=& {p(p^{2}-4m^{2})\over 4(p+q)}~. 
\label{eq:paraP}
\eea
The charge parameters $Q$, $P$ were used already in writing the 
solution above. Note that $q,p\geq 2m$, with equality corresponding
to the absence of electric or magnetic charge, respectively.

\subsection{Relation to Previous Work}
The black holes presented above are related to many different families
of black holes considered in previous work. The non-rotating black 
holes were found by Gibbons and Wiltshire~\cite{Gibbons:1986ac}.
They appear in a string theory context in~\cite{Cvetic:1995sz,Dhar:1998ip}.

The purely electric, or the purely magnetic, rotating 
black holes are special cases of the string theory black holes considered 
in~\cite{Horne:1992zy,Sen:1995eb}. Rotating black holes with
both magnetic and electric charges, but with respect to different
$U(1)$ fields, were found in a string theory context 
in~\cite{Cvetic:1996kv,Sen:1995eb,Jatkar:1996kd}. Sen~\cite{Sen:1995eb} 
further describes the construction of a very general black hole, including 
the ones considered here, but found it unpractical to carry out the details 
in full generality. Other strategies, which could lead in principle 
to the class of black holes considered here, were described 
in~\cite{Bakas:1994jc,Galtsov:1994pd}.

The ``diagonal'' case where electric and magnetic charges are equal $P=Q$, 
and rotation is absent, is the standard Reissner-Nordstr\"{o}m 
solution in Einstein-Maxwell theory. In~\cite{Khuri:1996xq,Sheinblatt:1998nt} 
it was found as a string theory solution, in this form. This is 
possible,
despite the coupling between the gauge field and the dilaton in the Lagrangean 
(\ref{eq:lag}), because the dual field strength has the inverse coupling so 
that, when the electric and magnetic charges are equal, it is consistent to 
take a constant dilaton in four dimensions. The more general solution found 
here shows that this phenomenon does not generalize to the rotating case: the 
standard charged, rotating Kerr-Newman black hole is {\it not} a special case 
of the family constructed here; in particular, it is not the diagonal case 
$P=Q$. 

The truncated theory defined by the Lagrangian (\ref{eq:lag}) is not
closed under electric-magnetic duality. One may attempt to make the
Lagrangian duality invariant by including additional fields, for example 
the axion. The derivation of the solution given above shows that it is 
consistent to set these fields to zero. However, doing so fixes the duality 
orbit so duality does not act as a solution generating transformation within 
the family constructed here. The electric and magnetic charges therefore 
appear as genuinely independent parameters in the metric. Other ways to 
obtain electric-magnetic duality involve the introduction of further gauge 
fields; for examples, see the literature discussed above.

\section{Black Holes in Four Dimensions}
\label{bh}
After dimensional reduction the five dimensional black string becomes
a four dimensional black hole with the metric:
\be
ds^{2}_{4,E} = - {H_{3}\over\sqrt{H_{1}H_{2}}}(dt + {\bf B})^{2}
+ \sqrt{H_{1}H_{2}}\left( {dr^{2}\over\Delta} + d\theta^{2} + 
{\Delta\over H_{3}}\sin^{2}\theta d\phi^{2}\right)~.
\ee
The matter fields are the gauge field ${\bf A}$ given in 
(\ref{eq:agauge}), and the dilaton:
\be
e^{-2\Phi_{4}}= \sqrt{H_{2}\over H_{1}}~.
\ee

The black hole has inner and outer horizons at $\Delta=0$, {\it i.e.}:
\be
r_{\pm} = m \pm \sqrt{m^{2}-a^{2}}~.
\label{eq:horizons}
\ee
The outer horizon is surrounded by an ergosphere, and shields a ring 
singularity. These and other features of the causal structure are
similar to those of Kerr black holes, see 
{\it e.g.}~\cite{Wald:1984rg}. The equations describing 
them are even the same, when written in terms of the parametric
variables $m,a$.

\subsection{Duality and other Symmetries}
After embedding into string theory the solutions are related to 
others by duality. As an example, consider the embedding into type II
string theory, compactified to four dimensions on $T^{6}$. Then the 
electric and magnetic charges can be interpreted as the KK-momentum 
and KK-monopole around any of the compact dimensions. 
Alternatively, as mentioned in the introduction, they are interpreted
as the D0-D6 system, or any of its obvious dual pairs of D-branes.

There are much more general possibilities. The subgroup of the general 
$E_{7(7)}$ duality group leaving the asymptotic Minkowski space invariant 
is the maximal compact subgroup $SU(8)$. A general strategy to investigate the 
orbit of these transformations is to consider the central charge matrix 
${\cal Z}_{AB}$ of $N=8$ supergravity in four dimensions, transforming 
as an antisymmetric tensor under $SU(8)$. From general formulae, {\it e.g.}
in~\cite{Balasubramanian:1998az}, it follows that in the present case 
the central charge matrix is particularly simple, of the form: 
\be
{\cal Z}=\lambda~{\rm diag}[\epsilon,\epsilon,\epsilon,\epsilon]~,
\ee
where $\lambda = P + iQ$ and:
\be
\epsilon = i\sigma_{2}= \pmatrix{ 0 & 1 \cr
                      -1 & 0 }~.
\ee
The four skew-eigenvalues of the central charge matrix are thus 
identical and equal to $\lambda$. $SU(8)$ duality transformations 
can at most change the skew-eigenvalues by phases. Moreover, 
the overall phase must be left invariant, because the 
duality group is $SU(8)$, rather than $U(8)$. It is therefore clear that, 
for generic $Q,P$, at least some of the skew-eigenvalues of the central 
charge matrix remain complex throughout the duality orbit.

The canonical family of black holes considered in the string theory 
literature depends on four independent 
charges~\cite{Cvetic:1996uj,Horowitz:1996ac}. 
The charges can be identified up to duality as the skew-eigenvalues of the 
central charge matrix. The eigenvalues of the central charge matrix
can thus be chosen real for the standard black holes; so these are 
distinct from the ones considered here, even after general dualities are 
taken into account.

The Einstein-Maxwell theory has a $U(1)$ duality symmetry acting on
the field strength. In view of the various dualities in string 
theory it is natural to expect an analogous symmetry in the present 
context. Such a symmetry, exchanging electric and magnetic charges in a 
continuous fashion, would multiply the four skew-eigenvalues by an 
identical phase. In general, this is evidently not a symmetry, since 
the phases do not multiply to unity. Thus there are no continuous 
duality symmetries that act within the family of solutions considered in 
this paper. The significance of this result is that the parameters $p$ 
and $q$ are genuinely independent.

Acting on the skew-eigenvalues with a phase is not generally a symmetry, 
but $i^{4}=1$ so this specific value generates a discrete duality
symmetry, interchanging the electric and magnetic charges. More 
precisely it takes $Q\to P$, $P\to -Q$, leaving the four-dimensional 
geometry invariant. A more explicit route to this symmetry starts from the 
truncated Lagrangean (\ref{eq:lag}), rewriting the Maxwell 
field in dual variables. This computation also shows that 
$\Phi_{4}\to -\Phi_{4}$ under the discrete duality.

Now, let us inspect the discrete symmetries of the solutions. One
symmetry has the parametric form:
\be
p\leftrightarrow q~~~;~~ a\to -a~.
\ee
It interchanges the functions $H_{1}\leftrightarrow H_{2}$,
inverting the dilaton $\Phi_{4}\to -\Phi_{4}$; and it leaves the four 
dimensional geometry invariant, except for ${\bf B}\to -{\bf B}$.
It also interchanges the electric potential $A_{t}$ with the magnetic 
potential $\Psi^{(2)}$ appearing in intermediate stages of the explicit
computations in section (\ref{comp}).

The discrete symmetry under time-reversal is superficially similar,
yet different in nature. It leaves the four dimensional geometry
invariant except for ${\bf B}\to -{\bf B}$; again, this takes
the angular momentum $J\to -J$. It also inverts the electric
potential $A_{t}\to -A_{t}$, but not the magnetic one $A_{\phi}\to 
A_{\phi}$. This implies $Q\to -Q$, $P\to P$.

The product of the two discrete symmetries described above is also 
a symmetry, taking $Q\to P$, $P\to -Q$, leaving $J$ invariant. It is 
this combined symmetry which is the discrete duality.

For another discussion of duality, emphasizing non-supersymmetric orbits, 
see~\cite{Ortin:1996bz}.

\subsection{The Extremal Limit and Stability}
\label{stability}
For a given value of the conserved charges $Q,P,J$, the lowest 
possible value of the mass occurs for $p,q\gg m,a$. This three 
parameter family of solutions is referred to as extremal.
In the extremal limit the parametric relations 
(\ref{eq:paraM}-\ref{eq:paraP}) can be inverted with the result:
\bea
2G_{4}M &=& (Q^{2/3}+ P^{2/3})^{3/2}~, \\
{a\over m} &=& {G_{4} J \over PQ}~.
\label{eq:aom} 
\eea
Note that the mass is independent of angular momentum; but nontrivial
dependence on the ratio (\ref{eq:aom}) can nevertheless be retained in 
the extremal geometry.

A natural benchmark for the mass is the BPS inequality, written in the 
present units as:
\be
2G_{4}M \geq \sqrt{Q^{2}+P^{2}}~.
\ee
This condition is always satisfied and cannot be saturated, when both
electric and magnetic charges are present. The extremal black holes can 
therefore not be supersymmetric.

A stronger benchmark is the comparison with the energy of two widely 
separated fragments, each carrying either the electric or the magnetic 
charge:
\be
2G_{4}M \geq Q+P~.
\ee
This inequality is also satisfied for all ranges of parameters. This 
shows that spontaneous fragmentation of the black hole into two parts 
is consistent with energy conservation. However, as discussed in the
introduction, the two constituents in the final state must satisfy 
Dirac's bound on the angular momentum:
\be
J \geq {PQ\over G_{4}}~.
\label{eq:dirac}
\ee
It is therefore only rapidly spinning black holes that are unstable 
towards this decay.

There are many other potential fragmentations, and angular momentum 
conservation may not generally rule such decays out. Fragmentation in 
identical lumps, each with charge assignments $(Q/2,P/2)$, is particularly
worrisome, because two identically charged fragments can have vanishing 
angular momentum. A nonrotating black hole with mutually prime quantized 
charges is not subject to this concern, but others are. It is 
nevertheless interesting that the most obvious decay channel is 
forbidden.

Independently of these considerations a large degree of stability is 
expected on entropy grounds. As discussed below, the black holes 
have considerable entropy, even in the extremal limit, and any
fragmentation would significantly lower the total entropy of the
system. 

The extremal family of solutions allow arbitrary value of $m/a\le 1$, 
so the inner and outer horizons (\ref{eq:horizons}) do not in general 
coincide. A computation of the inner and outer horizon area, given 
below, finds that those do agree, so there remains a sense that the horizons 
are ``close'' in the extremal case. In general relativity, the term extremality 
usually refers to a specific property of the causal structure, the appearance 
of a bifurcate Killing horizon. This is not the terminology used here.
It seems to require $r_{+}=r_{-}$, and so $m=a$. This condition is satisfied 
by a class of solutions parametrized by $q,p,m$, with no further constraints 
(beyond $q,p\ge 2m$). It would clearly be interesting to study the causal 
structure in more detail, and particularly to investigate the extremal limit 
closer. 

\subsection{Thermodynamics}
We next turn to the evaluation of the thermodynamical variables of the
black hole. 

The area of the black hole is determined from the four dimensional 
Einstein metric as:
\be
A = \int \sqrt{g_{\theta\theta}g_{\phi\phi}}d\theta d\phi 
= \int \sqrt{-H_{3}B_{\phi}^{2}} d\theta d\phi~.
\ee
This gives the black hole entropy:
\bea
S &=& {A\over 4G_{4}}={\pi\sqrt{pq}\over G_{4}} 
\left[ m + {pq+4m^{2}\over 2m(p+q)}\sqrt{m^{2}-a^{2}} \right] 
\nonumber \\
&=& 2\pi \left[ {m\sqrt{pq}\over 2G_{4}} + \sqrt{ {pq\over 16G_{4}^{2}}
({pq+4m^{2}\over p+q})^{2}-J^{2}} \right]~.
\eea
In the extremal limit the entropy simplifies to:
\be
S = 2\pi\sqrt{ {P^{2}Q^{2}\over G_{4}^{2}} - J^{2}}~.
\label{eq:extent}
\ee
Dirac's bound on the angular momentum (\ref{eq:dirac}) is precisely 
such that it forces the expression under the square root less than
or equal to zero. Inspecting the solution, this corresponds to a
singularity in the geometry outside the event horizon. Such solutions
are usually discarded. The remaining regular black holes are precisely 
the ones that are stabilized by angular momentum conservation. 

The limit $m\to a$ leads to the bizarre entropy formula:
\be
S = {4\pi m(p+q)\over pq+4m^{2}}J~,
\ee
which does not simplify further.

It is simplest to compute the inverse temperature at $\theta=0$, where the event 
horizon and the ergosphere meet. Here $g_{tt}=-g_{rr}^{-1}$ and:
\be
\beta_{H}={4\pi\over |g_{tt}^{\prime}|}~,
\ee
where the prime denotes differentiation with respect to the 
coordinate $r$. The result is:
\be
\beta_{H}=
{\pi\sqrt{pq}\over m}
\left[{pq+4m^{2}\over p+q}+{2m^{2}\over\sqrt{m^{2}-a^{2}}}\right]~.
\ee
The Hawking temperature vanishes in the extremal limit $p,q\gg m,a$,
and also in the limit $m\to a$. 

The rotational velocity of the black hole horizon is also determined
at $\theta=0$ where:
\be
\Omega_{H} = - {g_{\phi t}\over g_{tt}}= - {1\over B_{\phi}}~.
\ee
It can be written as:
\be
\Omega_{H}={2m(p+q)\over\sqrt{pq}\left( 
2m^{2}(p+q)+(pq+4m^{2})\sqrt{m^{2}-a^{2}}\right)}~a~.
\ee
Note that the product of the inverse temperature and the rotational velocity 
is particularly simple:
\be
\beta_{H}\Omega_{H} = {2\pi a\over\sqrt{m^{2}-a^{2}}}~.
\ee

It can be verified that the first law of black hole 
thermodynamics:
\be
dM = T_{H}dS + \Phi_{E}dQ + \Phi_{M}dP + \Omega dJ~,
\ee
is consistent with the results above. This computation also gives the
potentials conjugate to electric and magnetic charge:
\bea
\beta_{H}\Phi_{E} &=& {\pi\over 2mG_{4}}\sqrt{p(q^{2}-4m^{2})\over p+q}~
(p+{2m^{2}\over\sqrt{m^{2}-a^{2}}}) ~,\\
\beta_{H}\Phi_{M} &=& {\pi\over 2mG_{4}}\sqrt{q(p^{2}-4m^{2})\over p+q}~
(q+{2m^{2}\over\sqrt{m^{2}-a^{2}}}) ~.
\eea
Finally, homogeneity of the thermodynamic potentials gives the
generalized Smarr formula:
\be
2S = \beta_{H}M -\beta_{H}\Phi_{E} Q -\beta_{H}\Phi_{M} P -2\beta_{H}\Omega J~.
\ee
It is a good check on the computations that this formula is satisfied.

\subsection{Quantization Rules}
The electric and magnetic charges are quantized according to:
\bea
Q &=& 2G_{4}M_{0}~n_{Q}~, \\
P &=& 2G_{4}M_{6}~n_{P}~, 
\eea
where $n_{Q}$ and $n_{P}$ are integral. 
The mass parameters satisfy $8G_{4}M_{0}M_{6}=1$ so that:
\be
{2PQ\over G_{4}}= n_{Q}n_{P}~.
\label{eq:dquant}
\ee
The discussion in this paper uses (\ref{eq:dquant}), essentially Dirac's 
quantization rule, several times. However, the precise quantization rules
on the independent charges are less important. It may nevertheless be
useful to note that, if the electric and magnetic objects are interpreted 
in string theory as $D0$- and $D6$-branes, the masses are:
\bea
M_{0}&=& {1\over l_{s}g_{s}}~,\\
M_{6}&=& {V_{6}\over (2\pi)^{6}l_{s}^{7}g_{s}}~,
\eea
where the string units are defined so $l_{s}=\sqrt{\alpha^{\prime}}$ and 
$V_{6}$ is the volume of the six compact dimensions wrapped by the 
$D6$-brane. 

The extremal entropy (\ref{eq:extent}) can be rewritten as:
\be
S = 2\pi\sqrt{ {n_{P}^{2}n_{Q}^{2}\over 4} - J^{2}}~,
\ee
using the quantization rule (\ref{eq:dquant}). It is interesting that this 
expression involves only the quantized charges, the moduli cancel out. 
This is sometimes interpreted as a signal that a clear connection to the 
microscopic theory is possible~\cite{Larsen:1996ss}. For supersymmetric
black holes, the cancellation of moduli is understood as a consequence of 
enhanced supersymmetry at the horizon~\cite{Ferrara:1996dd}. However, 
this result does not seem to apply in the present circumstances.

Taking the cancellation of moduli into account, the duality group is
enlarged to the full noncompact group. In $N=8$ supergravity the extremal 
entropy formula for the full orbit is therefore:
\be
S = 2\pi\sqrt{{1\over 4}J_{4} - J^{2}}~,
\ee
where the quartic invariant of $E_{7(7)}$ satisfies $J_{4}>0$. The 
corresponding formula for the orbit of extremal black holes with a 
supersymmetric limit is:
\be
S =2\pi\sqrt{J^{2}-{1\over 4}J_{4}}~,
\ee
with $J_{4}<0$. The analogous entropy formulae for the $N=4$ theory are 
identical, except that the $J_{4}$ is replaced by the quartic S- and 
T- duality invariant.

\subsection{Comments on the Nonextreme Entropy}
The family of black holes considered here are all far from 
extremality, and so it is difficult to establish a firm connection with 
microscopic ideas for the entropy. The extremal (and diagonal) case was 
interpreted using the correspondence principle in~\cite{Sheinblatt:1998nt}; 
for a brane-antibrane interpretation of the general nonextreme case 
see~\cite{Dhar:1998ip}.

The purpose of this section is to apply the observations 
of~\cite{Larsen:1997ge,Cvetic:1997uw} in the present case. 
The idea is to interpret the black hole as a state in some underlying 
conformal field theory, with the effective levels of the state as the 
quantities that control the thermodynamics. Now, the proposal gives a 
{\it geometric} determination of the levels as:
\be
S^{(\pm)} = 2\pi ( \sqrt{N_{L}}\pm\sqrt{N_{R}})~,
\label{eq:spm}
\ee
where $S^{(+)}$ is the usual entropy, computed from the area of the event
horizon, and $S^{(-)}$ is computed similarly, but from the area of the
{\it inner} horizon. In the present case:
\be
S^{(-)} = 2\pi \left[ \sqrt{ {pq\over 16G_{4}^{2}}
({pq+4m^{2}\over p+q})^{2}-J^{2}}-
{m\sqrt{pq}\over 2G_{4}} 
~\right]~,
\ee
so the prescription leads to the levels:
\bea
N_{L}&=& {pq\over 16G_{4}^{2}}\left( {pq+4m^{2}\over p+q}\right)^{2} 
-J^{2}~,\\
N_{R}&=& {m^{2}pq\over 4G_{4}^{2}}~.
\eea
According to (\ref{eq:spm}) we have $N_{L}>N_{R}$ by convention. In
black holes with a supersymmetric limit this implies that the right 
movers are supersymmetric, but not the left movers. In the present 
context there is no supersymmetric limit; here the convention implies that
the left movers are supersymmetric, but not the right movers.

In general, it is not known what conformal field theory has these
expressions as effective levels and no independent verification is 
possible. However, general principles of conformal field theories 
nevertheless suffice for some consistency checks. An important requirement 
is that the difference of the levels must be an integer, by modular invariance. 
In the present case a simple computation gives:
\be
N_{L}-N_{R} = {Q^{2}P^{2}\over G^{2}_{4}}- J^{2} = 
{1\over 4}n^{2}_{Q}n^{2}_{P} -J^{2}~.
\ee
This is indeed an integer, for all values of the charges and the angular 
momentum. Note that this result is quite delicate: $J$ may be half-integer 
and $n_{Q}n_{P}$ may be odd, but the angular momentum quantization rules
for dyons ensure that these possibilities are correlated so that neither 
happen, or both happen at once. This is precisely what is needed.

In conclusion, we find that $N_{R}-N_{L}$ is integral, confirming a rule 
noted in many previous examples. This may be an indiction that both $N_{R}$ 
{\it and} $N_{L}$ are levels in some conformal field theory.

\subsection{The Magnetic Dipole Moment}
The magnetic dipole moment is a good indicator of the interplay between
the angular momentum and the electric and magnetic charges. It can also
be useful in establishing the connection to various microscopic 
models, see {\it e.g.}~\cite{Ferrara:1992nm,Duff:1997bs,Duff:1998ef}. 

The magnetic dipole moment can be
read off from the gauge potential ${\bf A}$, given in (\ref{eq:agauge}). 
The coefficient is determined by the precise analogy with the relation 
between angular momentum and the effective gauge potential ${\bf B}$. This 
procedure gives:
\be
\mu = \sqrt{p^{3}(q^{2}-4m^{2})\over (p+q)}~{a\over 4m}~.
\ee
A natural benchmark for the electric dipole moment is the classical
value:
\be
\mu_{0} = {QJ\over 2M} = \sqrt{p(q^{2}-4m^{2})\over 
(p+q)^{5}}~(pq+4m)q~{a\over 4m}~.
\label{eq:gyrb}
\ee
The gyromagnetic ratio $g$, defined through $\mu = g \mu_{0}$, becomes:
\be
g = {p(p+q)^{2}\over q(pq+4m^{2})}~.
\ee
To interpret this formula, consider some special cases.

The purely electric black hole has $p=2m$ and so the gyromagnetic 
ratio:
\be
g = {q+2m\over q}~.
\ee
This agrees with previous results~\cite{Horne:1992zy,Sen:1995eb}. 
The function decreases monotonically from Dirac's quantum value $g=2$, 
in the limit of vanishing electric charge ($q=2m)$, to the classical 
value $g=1$, for very large charge. The large charge limit is extreme 
and the result can be compared succesfully with our microscopic 
understanding~\cite{Duff:1998ef}. 
The appearance of Dirac's value in a natural limit of 
parameter space is intriguing, and reminiscent of the fact that this also 
happens for Kerr-Newman black holes. As in that case, it is not clear why 
Dirac's value should appear. 

A more illuminating special case may be that of vanishing electric 
charge $p=2m$, giving:
\be
g = {p\over 2m}({p\over 2m}+1)~.
\ee
The gyromagnetic ratio is always larger that Dirac's value $g=2$, 
obtained in the limit of vanishing magnetic charge $q=2m$.  It increases 
monotonically with the magnetic charge, with no upper bound. Thus,
the rotating {\it magnetic} background is very efficient at creating
an {\it magnetic} dipole moment. This property suggests some sort of phase 
separation of the microscopic constituents; however, the details are
puzzling.

As a last special case, consider the general extreme limit $p,q\gg m$
where:
\be
g = {(q+p)^{2}\over q^{2}} = \left[1 + \left({P\over 
Q}\right)^{2/3}\right]^{2}
\label{extg}
\ee
This attains the classical value for purely electric black 
holes, and increases without bound as the magnetic charge is 
turned on and becomes large. Similar qualitative behavior was noticed 
in~\cite{Jatkar:1996kd}, but for black holes that are not obviously
related to the present ones. The full function (\ref{extg})
characterizes properties of extremal black holes, and may be a good 
target for a microscopic analysis. 

The consideration of the electric dipole moment is completely parallel
to that of the magnetic dipole moment. Here one starts with the magnetic 
potentials, already computed in section (\ref{comp}). The resulting 
expressions for the magnetic variables can be found from 
the corresponding electric ones, {\it via} the interchange of variables 
$p\leftrightarrow q$. The associated discussion is therefore analogous.

\subsection{The Near-horizon Geometry}
In some cases the AdS/CFT correspondence~\cite{adsrev} offers a direct 
avenue from the near-horizon geometry of black holes to aspects of the 
underlying microscopic theory. As a first step in this strategy, consider 
the nonrotating black holes $a=0$, and take the limit $p,q\gg r,m$. The 
metric becomes:
\be
ds^{2}_{5} = {q\over p}dy^{2} - {2(p+q)\over q^{2}p}~(r^{2}-2mr)dt^{2}
+ {p^{2}q\over 2(p+q)}({dr^{2}\over r^{2}-2mr}+ d\Omega_{2}^{2})~.
\ee
The radius of the compact dimension is constant, so the geometry is
effectively four-dimensional. In fact, for $r\gg m$ it is precisely
$AdS_{2}\times S^{2}$, the near-horizon limit of extreme 
Reissner-Nordstr\"{o}m black holes. A general value of $m$ 
parametrizes a deformation away from the $AdS_{2}$ limit which 
preserves the asymptotic behavior. These results are promising, but
the conformal quantum mechanics dual to $AdS_{2}$ spaces is not well 
understood~\cite{Strominger:1998yg}, and the decoupling limit which defines 
it as a theory without gravity is more subtle than for other 
$AdS$-spaces~\cite{Maldacena:1999uz}.

Another decoupling limit for the $D0/D6$ system was considered 
in~\cite{Itzhaki:1998ka}. In this limit the effective near-horizon geometry 
is the five-dimensional Kerr black hole. The precise relation with the
$AdS_{2}\times S^{2}$ geometry is not clear.

When the rotational parameter is included the decoupling limit 
has $p,q\gg r,m,a$. In this limit the geometry is (\ref{eq:string}) 
with:
\bea
H_{1} &=& {p^{2}q\over 2(p+q)}\left( 1 - {a\cos\theta\over m} 
\right)~,\\
H_{2} &=& {q^{2}p\over 2(p+q)}\left( 1 + {a\cos\theta\over m} \right)~,\\
B_{t} &=& {\sqrt{p^{3}q^{3}}\over 2(p+q)}~(r-m)H_{3}^{-1}
~{a\over 2m}\sin^{2}\theta ~,\\
A_{t} &=& - \sqrt{p+q\over q}~, \\
A_{\phi} &=& -2P~{m\cos\theta + a\over m+ a\cos\theta}~,
\eea
and $H_{3},\Delta$ retained in full. The parameter $a/m$ must be less 
than unity for the geometry to be regular, in agreement with the discussion 
in section~\ref{stability}. The scale of the compact dimension is set by 
$H_{2}/H_{1}$ and is independent of $r$, but dependent of $\theta$. Thus 
the compact direction effectively becomes part of the angular $S^{2}$.

The azimuthal coordinate mixes with the time coordinate, as expected for 
rotating spacetimes. However, $G_{\phi,t}\sim r$ for large $r$ so the 
$AdS_{2}\times S^{2}$ geometry is not recovered asymptotically. 
To interpret this result, recall that rotation similarly modifies black 
holes with near-horizon geometry $AdS_{3}\times S^{2}$ by mixing the $AdS_{3}$ 
with the $S^{2}$, but in this case the asymptotic geometry is unaffected by the 
rotation~\cite{Cvetic:1999ja}. The field theory interpretation is that the 
rotational parameter characterizes a specific excitation, rather than a 
deformation of the theory. In the present case we reach the opposite 
conclusion: the deformation due to rotation is so large that it 
modifies the dual theory.

One may take further limits between the small parameters, in an attempt
to find the geometry corresponding to the vacuum of the dual CFT. The
relation $r\gg m,a$ is natural because it retains the ratio $a/m$
characterizing the rotation. However, no special simplication seems to
occur in the limit.

{\bf Acknowledgments:}
I thank G. Horowitz, M. Krogh, R. Myers, A. Peet, and other participants 
in the SUSY99 program at the ITP for discussions. This work was supported 
by DOE grant DE-FG02-90ER-40560 and by a Robert R. McCormick Fellowship. I 
thank the theory group at the University of Michigan in Ann Arbor and the 
ITP at the University of California in Santa Barbara for hospitality.

{\bf Note Added:}
The black hole solutions presented in this paper were previously found by
Rasheed~\cite{Rasheed:1995zv}; and independently by Matos and 
Mora~\cite{Matos:1996km}. The discussion of the solution in the present
work is substantially new. I thank C-M. Chen and T. Matos for bringing the 
earlier works to my attention.
 

\end{document}